  \documentclass[journal,twocolumn]{IEEEtran}

\usepackage[colorlinks,urlcolor=blue,linkcolor=blue,citecolor=blue]{hyperref}

\usepackage{amsmath}
\usepackage{amssymb}
\usepackage{graphicx}
\usepackage{url}
\usepackage{cite}
\usepackage{mathtools}
\usepackage{cuted}
\usepackage{etoolbox}
\usepackage{subcaption}
\usepackage{tikz}
 \usepackage{diagbox}
\usepackage{multirow}
\let\labelindent\relax
\usepackage{enumitem}
 \newcolumntype{P}[1]{>{\centering\arraybackslash}p{#1}}

\newcommand{\CL}{\color{blue}}
\newcommand{\CB}{\color{black}}
\usepackage{fourier} 
\usepackage{array}
\usepackage{makecell}

\usepackage{pifont}
\newcommand{\cmark}{\ding{51}}
 
\begin{document}

	\title{Evolution Toward 6G Wireless Networks: A Resource Management Perspective}
	\author{Mehdi Rasti, Shiva Kazemi Taskou, Hina Tabassum, and Ekram Hossain}

	\maketitle	
	\begin{abstract}
	In this article, we first present the vision, key performance indicators, key enabling techniques (KETs), and services of 6G wireless networks. Then, we highlight a series of general resource management (RM) challenges as well as unique RM challenges corresponding to each KET.  The unique RM challenges in 6G necessitate the transformation of existing optimization-based solutions to artificial intelligence/machine learning-empowered solutions. In the sequel, we formulate a joint network selection and subchannel allocation problem for 6G \textit{multi-band} network that provides both further enhanced mobile broadband (FeMBB) and extreme ultra reliable low latency communication (eURLLC) services to the terrestrial and   aerial   users. Our solution highlights the efficacy of multi-band network and demonstrates the robustness of dueling deep Q-learning in obtaining efficient RM solution with faster convergence rate compared to deep-Q network and double deep Q-network algorithms. 
	
	
	\end{abstract}

\begin{IEEEkeywords}
 6G, resource management, further enhanced mobile broadband (FeMBB), extreme ultra reliable low latency communication (eURLLC), artificial intelligence, machine learning, deep reinforcement learning   
\end{IEEEkeywords} 	
	
	\section{Introduction}
	The emergence of  a new generation of wireless networks  becomes inexorable every decade since early 1980s. 
Currently, 5G-NR has a multitude of advantages over the long-term evolution (LTE)/LTE-advanced technology, i.e. higher data rates ($\sim$0.1Gbps), low latency  ($\sim$1-10msec), higher mobility ($\sim$500km/h), and support to 10$^6$ devices$\mathrm{/km^2}$. According to International Telecommunication Union (ITU), three predominant use cases of 5G include  ultra reliable low latency communication (uRLLC), enhanced mobile broadband (eMBB), and massive machine type communication (mMTC) \cite{ITU} that leverage   millimeter-wave (mm-wave) communication, large-scale antenna arrays (i.e. massive MIMO), and ultra-dense deployment of access points.
	
	Despite the advancements in 5G, the evolving smart infrastructure, efficient technologies, and diversified wireless applications (e.g. connected and autonomous vehicles, virtual and augmented reality, remote surgery and holographic projection) make the launch  of  sixth  generation (6G) networks inevitable. Different from 5G, 6G  networks are envisaged as \textit{multi-band, decentralized, fully autonomous}, and \textit{hyper-flexible  user-centric} systems encompassing satellite, aerial, terrestrial, underwater, and underground communications. 

\begin{figure*}
    \centering
    \includegraphics[scale=0.17, trim=4 4 4 4 ,clip]{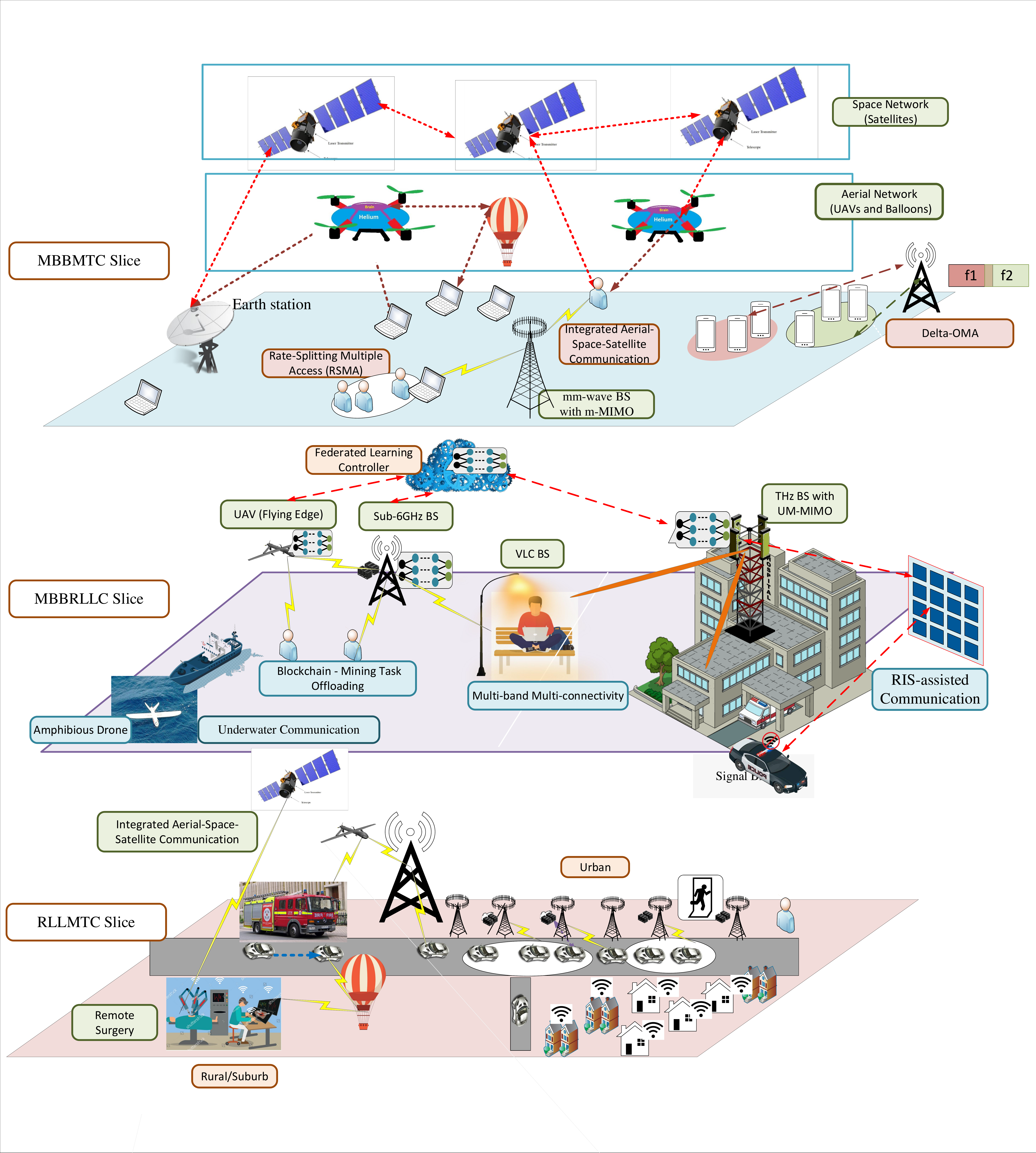}
    \caption{Vision of intelligent 6G network architecture  with multi-band aerial/terrestrial/satellite/underwater connectivity.}
    \label{fig:my_label}
\end{figure*}

In this article, we first present the vision of 6G, the key performance indicators (KPIs), as well as the  key enabling techniques (KETs), and services. Then, we clarify how 6G KETs alter the perspective of resource management (RM) by extending the set of decision variables, constraints, and performance measures. We then pin-point the challenges offered by each KET.  The RM challenges necessitate the transformation of existing solutions based on traditional optimization theory. In the sequel, we formulate a \textit{joint network selection and subchannel allocation (JNSA)} problem for 6G \textit{multi-band} networks (MBN). We employ three deep reinforcement learning (DRL) methods, namely deep Q network (DQN), double DQN, and dueling DQN (DuelDQN) to address the JNSA problem. Our results show the efficacy of MBN model as well as DuelDQN method compared to other DRL methods in terms of convergence rate and its robustness. 

\section{6G: Vision, KPIs, KET,  and Services}

The KPIs include a peak data rate of up to  1Tbps and user-experienced data rate of up to 1Gbps, over the air latency of   10-100${\mu}$s, and end-to-end latency of  1ms; 99.9999999\% reliability; energy-efficiency of   1$\mathrm{Tb/J}$ with battery life expectancy of  20  years which is 10-100 times more than 5G; connectivity of up to   10$^7$ devices$\mathrm{/km^2}$; mobility of $1000\mathrm{km/h}$ and indoor and outdoor positioning of $  10 \mathrm{cm}$ and $  1 \mathrm{m}$, respectively \cite{mvt-2019}, \cite{flagship-oulu}. 
	
	To achieve the aforementioned KPIs, the physical (PHY) and medium access (MAC) layer of 6G will benefit from the  KETs that include
\textbf{(i)} terahertz (THz) and visible light communications (VLC) ranging from 0.1--10 THz (licensed) and  400--800  THz (unlicensed), respectively,
\textbf{(ii)} ultra-massive spatially modulated MIMO (UM-SM-MIMO),
\textbf{(iii)} reconfigurable intelligent surface (RIS), 
\textbf{(iv)} {U}nderwater-{T}errestrial-{A}ir-{S}pace integrated {Net}works (UTASNet), 
\textbf{(v)} dynamic network slicing, virtualization, and
\textbf{(vi)} variants of non-orthogonal multiple access (NOMA) including delta-OMA (D-OMA)and rate splitting multiple access (RSMA) \cite{6G-Tech}. 

In addition, 6G will not only be instrumental in the provisioning of enhanced 5G services (i.e. extreme ultra reliable low latency communication [eURLLC], further enhanced mobile broadband [FeMBB], ultra massive machine type communication [umMTC]), but also will attempt to unify  diversified 5G services. For instance, mobile broadband reliable low latency communication (MBBRLLC) consolidates FeMBB and eURLLC to support high broadband data rates along  with reliable, and low latency communication. Similarly,  mobile broadband machine type communication (MBBMTC) will unify  FeMBB and umMTC to support high broadband data rates along with massive connectivity, and reliable low latency machine type communication (RLLMTC)  will blend eURLLC and umMTC to support massive connectivity, reliability, and low latency.   Finally, mobile broadband and reliable 
low latency machine type communication (MBBRLLMTC) is proposed to amalgamate eURLLC, FeMBB, and umMTC to offer high data rates, reliability, low latency, and massive connectivity \cite{mvt-2019}. 

The vision of 6G is illustrated in Fig. \ref{fig:my_label}, while Table~\ref{requirement-technologies} summarizes how the 6G services and KETs satisfy the technical requirements of 6G.

\section{Resource Management Challenges in 6G}

			\begin{table*}
		\centering
		\caption{Technical requirements, services, and KETs for 6G }
		\label{requirement-technologies}
		\begin{tabular}{{|m{1cm}| m{2.cm}| m{1.75cm} | m{0.75cm}| m{0.45cm}| m{1.15cm}| m{1.15cm}| m{1.2cm}| m{1.3cm}| m{1cm}|}}
			\hline
			\multicolumn{2}{|c|}{\diagbox[width=15 em, height=5 em]{Services and KET(s)}{\\   Requirements}} &
			\centering Peak$ \backslash $user-experienced data rate & 
			\centering Latency & 
			\centering Jitter & 
			\centering Reliability & 
			\centering Energy efficiency &
			\centering Spectrum efficiency & 
			\centering Connectivity & 
			\centering\arraybackslash Mobility \\
			\hline
			\hline
			
			\centering\multirow{7}{*}{Services} & eURLLC &    &  \centering \cmark & \centering \cmark  & \centering \cmark &      &   &      &   \centering\arraybackslash\cmark   \\ 
			\cline{2-10}
			
			& FeMBB & \centering \cmark &      &      &      &      &  \centering \cmark &      &  \centering\arraybackslash\cmark \\				
			\cline{2-10}
			
			& umMTC &     &      &      &      & \centering \cmark &      & \centering \cmark  &    \\				
			\cline{2-10}
			
			& MBBRLLC & \centering \cmark &  \centering \cmark   & \centering \cmark  & \centering \cmark &   & \centering \cmark  &    & \centering\arraybackslash \cmark  \\				
			\cline{2-10}
			
			& MBBMTC & \centering \cmark &      &     &     &\centering \cmark  & \centering \cmark  & \centering \cmark  &  \centering\arraybackslash \cmark \\				
			\cline{2-10}
			
			& RLLMTC &    & \centering \cmark  & \centering \cmark & \centering \cmark & \centering \cmark &      &  \centering \cmark &      \\				
			\cline{2-10}
			
			& MBBRLLMTC & \centering \cmark & \centering \cmark   &  \centering \cmark   & \centering \cmark & \centering \cmark  & \centering \cmark  & \centering \cmark  & \centering\arraybackslash \cmark  \\				
			\hline
			\hline 
			
			\centering\multirow{7}{*}{ KET(s)} & THz& \centering \cmark  & \centering \cmark  &     &\centering \cmark &      &      &      &      \\ 
			\cline{2-10}
			
			& VLC &\centering \cmark  &      &     &     &\centering \cmark  &      &      &   \\ 
			\cline{2-10}
			
			& UM-SM-MIMO and RIS  &\centering \cmark  &\centering \cmark  & &\centering \cmark &\centering \cmark  &\centering \cmark  &\centering\arraybackslash \cmark &       \\ 
			\cline{2-10}
			
			& AI  &      &\centering \cmark &     &\centering \cmark  &      &      &      &       \\ 
			\cline{2-10}
			
			& UTASNet  &      &      &     &     &      & \centering \cmark &\centering \cmark  &\centering\arraybackslash \cmark   \\
			\cline{2-10}
			
			& Blockchain  & \centering \cmark &      &     &     &      &\centering \cmark  &      &      \\ 
			\cline{2-10}
			
			& Dynamic network slicing  &\centering \cmark  &\centering \cmark  &\centering \cmark &     &  \centering \cmark &      & \centering \cmark &       \\ 
			\cline{2-10}
			
			
			& RSMA &\centering\cmark  &\centering \cmark  &     &\centering \cmark &\centering \cmark  &\centering \cmark  &\centering\arraybackslash \cmark  &        \\ 
			\cline{2-10}
			
			
			
			\hline			
			
		\end{tabular}
	\end{table*}	
	
In this section, we systematically present the general and specific RM challenges in 6G networks. 
\subsection{General Challenges}
\begin{itemize}[leftmargin=*]
	\item \textit{Extended  set of decision variables}:
	With the integration of each KET,  a new decision variable(s) is added to the set of decision variables resulting in an increased degree of freedom and  flexibility to meet the KPIs. For example, phase-shift optimization in RIS networks, height optimization in aerial networks, opportunistic spectrum selection in mixed RF/mm-wave/THz networks, mining mode selection in blockchain networks, etc.  When two or more KETs coexist, the degree of freedom increases further resulting in a more cumbersome RM problem.
	\item \textit{Extended set of constraints:} With the emergence of smart user devices enabled with multi-connectivity  and upcoming 6G KETs, a variety of traditional constraints such as assigning a given subchannel to at most one user in a single cell, or associating a given user to a single  BS in both the uplink and downlink  will need to be modified or relaxed.   
For example, in a multi-band UTASNet, users can associate with two or more BSs  simultaneously. Also, due to new multiple access mechanisms,  new forms of interference will need to be incorporated. 
	
\item \textit{Novel KPIs:}
KPIs in 6G will focus on mobility-aware performance, quality of experience (QoE), volumetric spectral  efficiency (in bps/Hz/m$^3$) and energy efficiency,  resource efficiency, and degree of intelligence at devices.
QoE is defined as perceived end-to-end user’s experience and expectation.  
Also, in 6G,  quality of physical experience (QoPE)  will be considered which incorporates human physiological factors with quality of service (QoS) and QoE \cite{QoPE}.

\item \textit{Joint optimization:} Joint optimization approaches for RM, such as joint uplink and downlink resource allocation (e.g. in full-duplex networks), joint multi-cell resource allocation (e.g. subchannel and power allocation in multi-cell NOMA systems), joint RM for radio access network (RAN) and edge computing severs, and cross-layer optimization (e.g. joint optimization of edge caching and user BS association) can improve network performance in comparison with  disjoint optimizations. This is because, the joint optimization approaches expand the feasible solution of the problem and increase the degrees of freedom.  


\item { \textit{Data-driven RM:}}  Leveraging ML tools and large volumes of data, 6G will benefit from faster and real-time RM solutions without  explicit mathematical models. 
 In other words, data-driven RM with AI can potentially allocate resources dynamically according to the requirements, makes it possible to allocate resources with the knowledge extracted from big data without the need of explicit mathematical models, and enables operators making real-time decisions.  Finally, the performance criteria of 6G as shown in Table \ref{requirement-technologies} should be jointly considered with AI performance metrics such as prediction accuracy and convergence time.

\item { \textit{Convergence of communication, computing, caching, control, sensing,  and localization (4CSL)}:}  
6G will be a converged network of 4CSL functionalities. With this convergence, 6G will integrate the control systems with wireless communications in the cyber-physical systems and support Internet-of-everything applications such as augmented reality, mixed reality, and extended reality that  require localization, communication, computing, and caching  techniques/resources. Such convergence in 6G can be realized and accelerated with the help of AI. 
4CSL complicates the resource management in 6G, since with a massive number of devices/objects, communication, computing, and caching resources should be efficiently allocated. 
Alongside, system requirements such as stability and low-latency can be achieved with  efficient joint control techniques and RM.
\end{itemize}
The above factors systematically translate into the key RM challenges indicated in Table~\ref{decision-variables-technologies}, where the decision variables corresponding to individual and integrated KETs are presented. In what follows, we  explain the RM challenges corresponding to individual KET.


\subsection{Specific Challenges Corresponding to KETs}
	\subsubsection{Multi-band Networks}
		6G will operate on a variety of transmission frequencies including radio frequency (RF), mm-wave, VLC, and THz frequencies. The variety of spectrum offers a trade-off among coverage area, capacity, users' mobility, and latency. The coexistence of multiple frequencies can be realized in two ways, namely, \textit{coexisting} and \textit{hybrid} deployment. In the \textit{coexisting} approach, the BSs with different frequencies are deployed separately and each BS at each time can operate on only one of RF, THz, and VLC frequency bands. In the \textit{hybrid} approach, each BS can operate on more than one frequency bands. Optimizing the deployment of BSs, traffic-load aware network activation mechanisms,  opportunistic spectrum selection at the users' end, and multi-connectivity solutions  will be the primary  challenges. Furthermore, due to  significantly varying coverage zones, mobility-aware resource management will be more predominant than ever before.

		\subsubsection{UM-SM-MIMO}
		 Compared to the 5G in which $ 256-1024 $ antennas realize massive MIMO, 6G will deploy more than $ 10,000 $ antennas, i.e. ultra-massive spatially modulated MIMO (UM-SM-MIMO). In SM-MIMO, a  transmitter can transmit information over few antennas than conventional MIMO, thus minimizes active antenna elements. However, sub-optimal spectral efficiency, faster antenna switching, and training overhead make RM a challenging problem.
		 
\subsubsection{Metasurfaces (or Reconfigurable Intelligent Surfaces [RIS])} Metasurfaces consist of a series of reconfigurable elements that, when the direct connection between the user and the BS is weak, reflects/refracts/absorbs the signal  by altering the phase, amplitude, or frequency of the signal sent from the transmitter. Subsequently, optimizing the deployment intensity and locations of metasurfaces, phase-shifts of the elements along with the mode of each element make the  RM problem challenging.
	\begin{table*}
	\centering
	\caption{The decision variables and and RM challenges corresponding to individual  KETs in 6G  (CR allocation stands for common resource allocation, including user association, subchannel allocation, and transmit power) }
	\label{decision-variables-technologies}
	\begin{tabular}{{| m{2cm}| m{5 cm}| m{ 8 cm}|}}
		\hline
		KET & Decision Variables
		& RM Challenges/Problems
		\\
		\hline
		\hline
		Multi-band (RF/THz/VLC) communication & 
		\begin{itemize}[labelindent=0em,labelsep=0.11cm,leftmargin=*]
			\item CR allocation 
			\item Network  selection (selecting RF, THz, or VLC frequency bands)
			\item  Multi-connectivity degree
			\item  Users' field of view 
			\item Selection of VLC/THz access points mode (sleeping or active) 
			\vspace{-1 em}
		\end{itemize}
		& 
		\begin{itemize}[labelindent=0em,labelsep=0.11cm,leftmargin=*]
			\item Trade-off among coverage area, capacity, and latency 
			\item Providing service requirement of eURLLC, FeMBB, and umMTC 
			\item Load balancing 
			\item VLC/THz access point placement  
			\item Single or multi-connectivity 
			 \item  Dimming control  in the VLC link 			to guarantee both eye safety and satisfy practical illumination 			constraints 
			\vspace{-1 em} 
		\end{itemize}   \\ 
		\hline

		AI  &  
		\begin{itemize}[labelindent=0em,labelsep=0.11cm,leftmargin=*]
			\item CR allocation 
			\item Selection of participated users  (client scheduling)    
			\item Learning data selection
			\item Latency of one training period
			
			\vspace{-1 em} 
		\end{itemize}   
		&  
		\begin{itemize}[labelindent=0em,labelsep=0.11cm,leftmargin=*]
			\item  RM considering learning metrics, including loss, accuracy, and convergence time 
			\item Assuring latency and reliability of learning
			\item Finding the appropriate number of iterations 
			\item Joint uplink and downlink RM
			\vspace{-1 em} 
		\end{itemize} \\ 
		\hline
			UM-SM-MIMO and RIS  & 
		\begin{itemize}[labelindent=0em,labelsep=0.11cm,leftmargin=*]
			\item CR allocation 
			\item  RIS allocation 
			\item Reflection coefficient (e.g. phase shift)
			\item Active/passive RIS mode selection
			\item Beamforming
		
		\vspace{-1 em} 
		\end{itemize} 
		& 
		\begin{itemize}[labelindent=0em,labelsep=0.11cm,leftmargin=*]
			\item RIS placement 
			\item Beamforming management
		\end{itemize}\\
		\hline
	
		UTASNet  & 
		\begin{itemize}[labelindent=0em,labelsep=0.11cm,leftmargin=*]
			\item CR allocation  
			\item   UAVs' trajectory 
			\item  UAVs' location
			\item  UAVs' hovering time allocation 
			\vspace{-1 em} 
		\end{itemize}    
		&
		\begin{itemize}[labelindent=0em,labelsep=0.11cm,leftmargin=*]
			\item Intra and inter-network interference management
			\item  Satisfying service requirements of eURLLC, FeMBB, and umMTC
			\item Connection offloading and load balancing 
			\item Route and placement of moving BSs 
			\item Path-planning
			\vspace{-1 em} 
		\end{itemize}\\
		\hline
		
		Blockchain  & 
		\begin{itemize}[labelindent=0em,labelsep=0.11cm,leftmargin=*]
			\item CR allocation 
			\item Mining mode selection 
			\item Block size   
			
			\item  Block interval (or block arrival rate)
			\item  Block producer selection
			\item Full function node deployment
			
			\vspace{-1 em} 
		\end{itemize}
		& 
		\begin{itemize}[labelindent=0em,labelsep=0.11cm,leftmargin=*]
			\item  Optimizing orphaning probability, reward, transaction confirmation rate, and fork probability 
			 \item Designing appropriate consensus and incentive mechanisms 
			
		\end{itemize}    \\ 
		\hline	
		
		NOMA with its variants (RSMA, D-OMA, and P-NOMA) & 
		\begin{itemize}[labelindent=0em,labelsep=0.11cm,leftmargin=*]
			\item  CR allocation 
			\item  User clustering 
			\item Decoding order selection
			\item Common and private data rate 
			\item Users' common signal selection
			\vspace{-1 em} 
		\end{itemize}
		& 
		\begin{itemize}[labelindent=0em,labelsep=0.11cm,leftmargin=*]
			\item Intra and inter-cell interference management  
		\end{itemize}\\ 
		\hline
		
		Dynamic network slicing  & 
		\begin{itemize}[labelindent=0em,labelsep=0.11cm,leftmargin=*]
			\item CR allocation 
			\item Slice admission control 
			\item  User-slice association
			\vspace{-1 em}  
		\end{itemize}
		&
		\begin{itemize}[labelindent=0em,labelsep=0.11cm,leftmargin=*]
			\item  Slice isolation 
			\item Providing diverse service requirement of eURLLC, FeMBB, and umMTC slices
			\item Providing combined service requirements for some applications
			\vspace{-1 em} 
		\end{itemize}   \\ 
		\hline
	\end{tabular}
	
\end{table*}

	\subsubsection{UTASNet}	6G  will  be  an  integrated  network  of  {U}nderwater (or sea), {T}errestrial, {A}ir, and {S}pace {Net}works (UTASNet). UTASNet creates severe challenges for RM, due to heterogeneity in the transmitter/receiver, mobility, maximum transmit powers,  computation and storage capabilities, interference, and channel propagation characteristics. Furthermore, while UTASNet can lead to high throughput and unlimited wireless connectivity, providing eURLLC services in UTASNet is  challenging due to the large turn-around communication time in aerial, space, and underwater communications, channel coherence, and the mobility of BSs and users. For efficient deployment of UTASNet, optimal clustering of diverse BSs and user offloading schemes would be crucial. 

		 	\subsubsection{NOMA and its Variants}
			
		To enable massive connectivity in 6G, NOMA along with its variants (e.g. RSMA, D-OMA, P-NOMA, etc.) will be the new norm.  
		For instance, D-OMA~\cite{doma} exploits partial overlapping of adjacent subchannels that are assigned to different clusters of users served by NOMA. Thus, the performance of D-OMA critically depends on the number of users in  a NOMA cluster, the fraction of overlapping spectrum, and subchannel scheduling.  Partial overlapping of adjacent subchannels  can yield  severe  interference. Therefore, it is crucial to optimize the scheduling, NOMA cluster size, and the fraction of overlapping spectrum  efficiently. On the other hand, in RSMA, each user's data rate is equal to the sum of the data rate from the common signal and the private signal. Because a common signal is a combination of multiple users' signals, the data rate allocation to each user is a decision variable.		

	\subsubsection{Dynamic Network Slicing}
	6G is expected to provide different services with diverse requirements through network slicing. Due to the end-to-end nature of the QoS, network slicing should be done in an end-to-end manner from the RAN to the core and transport networks. Moreover, since service demands and network conditions vary dynamically, slices need to be dynamically created, modified, and deleted, which requires resources to be flexibly and dynamically allocated to logical networks based on their service requirement. 
	It is noteworthy that network slicing for provisioning of individual URLLC, eMBB, and mMTC services was studied in 5G. Nonetheless, providing 6G specific services, including eURLLC, FeMBB, umMTC, and the possible combination of them require new solutions. 
		
		 \subsection{Other RM Challenges}
		 \subsubsection{Application-Driven Cross-Layer RM} In 6G systems, cross-layer resource management may be needed considering application layer performance requirements.  As an example, in a blockchain-enabled wireless/mobile network, RM is  challenging, because in addition to traditional QoS provisioning in wireless networks, the performance metrics such as transaction confirmation rate, winning, orphaning, and forking probabilities should also be taken into account. 
		In a mobile blockchain network, the probability of becoming an orphan block depends on wireless communication characteristics and the offloading of the mining process on mobile edge computing  servers.
		Furthermore, in mobile blockchain networks, block size affects orphaning probability, reward, and transaction confirmation rate. 
Optimizing the block size and  mining mode selection (solo versus pool mining) is thus a challenging RM problem.
		
			\subsubsection{Machine Learning (ML)-Enabled RM}
		  ML will be one of the key techniques to be used for RM in 6G networks. The ML techniques can be centralized or distributed. In the former, users and BSs send their local data to a centralized entity which results in  communication overheads, network resource consumption,  privacy concerns, etc. 
		    On the other hand, distributed ML (e.g. federated learning) is challenging due to the limitation of local computation and energy  resources, and dynamics of wireless communications which affect the global aggregation procedure.  Developing an appropriate global model for  the non-identical data is a challenge. Furthermore, the neural network training, freshness of the data, and  labeling the training data are challenging issues.
		  
		   \subsubsection{Digital Twin (DT)} DT will enable efficient RM of scarce computing and communication resources. 
		  In DT, computing time and communication time are inversely related \cite{DT}, so the joint allocation of computing and communication resources should be done. In addition, for different services in 6G, there will be different DTs that may have conflicting aims and requirements.

	\section{ML-Enabled Resource Management for 6G}

The existence of a variety of  discrete decision variables in 6G, enumerated in Table \ref{decision-variables-technologies}, render most of the RM problems non-convex and non-linear mixed integer programming problems which cannot be well addressed by traditional optimization methods. 
Furthermore, employing traditional optimization methods with uncertain channel status information, traffic load, the number of users, and complex dynamics of wireless channels results in  non-scalable solutions. Recently, ML techniques have been shown to be computationally efficient with discrete decision variables,  high-dimensional feature space, and uncertainty in wireless networks, as demonstrated in what follows for an  instance  of  6G   multi-band wireless  networking  scenario  with  FeMBB  and  eURLLC  services.

\subsection{System Model and Problem Formulation}

We consider an MBN in which a radio base-station (RBS)  located in the cell-center  coexists with the THz base-stations (TBSs). The MBN provides both FeMBB and eURLLC services to  the terrestrial and aerial users randomly distributed in the coverage area.
We formulate JNSA problem to jointly maximize FeMBB users' data rate and eURLLC users' reliability. The QoS requirement of FeMBB users is guaranteed by the data rate constraint of each FeMBB user. FeMBB users' data rate is obtained through Shannon's capacity formula which depends on selected network and allocated subchannels. Also, for each eURLLC user, reliability requirement is defined so that the decoding error probability does not exceed a given maximum value. Based on the finite blocklength capacity model, eURLLC users' decoding error probability is \cite{uRLLC}:  
$$ \epsilon=Q\left(\sqrt{\dfrac{L_B}{V}}  \left(\log_2(1+\gamma)-\dfrac{DM}{T {w}}\right)\right), $$ where $ Q $ is Gaussian Q-function,  $ L_B $ is blocklength in symbols,  $ V $ is the channel dispersion obtained by $ V=1-\left({1+\gamma }\right)^{-2} $, $ \gamma $ is the SINR, $ D $ is the number of bits in each block,  $ T $ is the duration time to transmit a block,  $ w $ is the bandwidth of the subchannel, and $ M $ is the number of mini-slots in each subchannel.

To share the frequency bands between FeMBB and eURLLC users, we use the puncturing approach. In the puncturing approach defined by 3GPP \cite{3GPP-Puncturing}, the subchannels are assigned to FeMBB users, and some of the mini-slots in subchannels are assigned to eURLLC users. Due to puncturing, the FeMBB users' data rate is lost, i.e. the actual data rate of FeMBB users is equal to $$ R^{\mathrm{actual}}=w\log_2\left(1+\gamma\right)-\left(\dfrac{\sum_{i=1}^{N}\sum_{m=1}^{M}\beta_i^m}{M}\right)w\log_2\left(1+\gamma\right),$$  where   $ N $ is the number of eURLLC users and $ \beta_i^m $ is the mini-slot allocation variable to eURLLC users.

\subsection{Proposed Solution}
We convert the multi-objective JNSA problem to  a single-objective problem with the weighted sum method \cite{multi-objective}.  Due to the integer characteristic of network selection and subchannel allocation variables, we employ value-based DRL methods. 
We employ DRL in which each user is an agent. 
Through  DRL, each agent (user) learns which BS or subchannel to choose to ensure its QoS and maximize its objective function (reward). 
In DQN, which is a DRL-based approach, Q-values are  approximated by a deep neural network (DNN). When the action-state space is large, DQN may overestimate Q-values since it uses the same values to select and evaluate action. To overcome this drawback, the double DQN algorithm is employed. In DQN and double DQN, Q-values are estimated to select actions in all states, while in many states, the choice of action does not affect agents' learning performance. Hence, to measure the importance of each action, the DuelDQN algorithm is employed in which the output of the DNN is divided into two streams, one to estimate the value function and the other to describe the importance of each action over other possible actions. 
\begin{figure}
		\centering
		\includegraphics[width=0.45\textwidth]{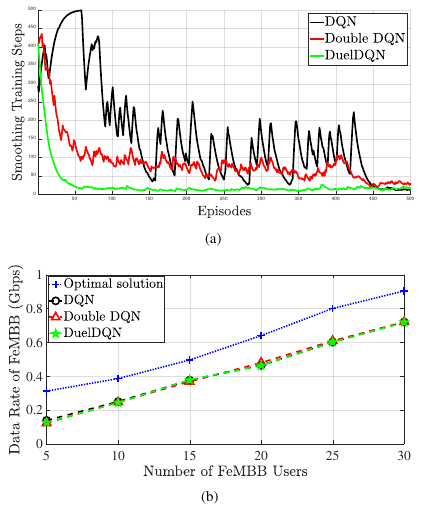}
		\caption{Comparison of DQN, double DQN, and DuelDQN algorithms in terms of (a) convergence and (b) performance.}
		\label{fig:simulation_1}
	\end{figure}

\subsection{Simulation Parameters}
We consider a coverage area of radius $ 500 $m in which an RBS is located in the center and $ 20 $ TBSs, each covering a radius of $ 5 $m, are distributed. The maximum power of RBS and TBSs are set to $ 10 \mathrm{W} $ and $ 1 \mathrm{W} $, respectively. Also, a number of terrestrial and aerial users are distributed in this area. The total bandwidth for  RF and THz is set to $ 20\mathrm{MHz} $ and $ 10\mathrm{GHz} $  which is divided into 20 orthogonal subchannels. 
%
The path-gain between user $ i $ and BS $ j $ at subchannel $ k $ for RF communication is obtained by $ h_{i,j}^k=\left({c}/{4\pi f_{\mathrm{RF}}}\right)^2 x^k d_{i,j}^{-\alpha} $, where  $c$ is the speed of light,
$ f_{\mathrm{RF}}=2.1\mathrm{GHz} $ is the RF carrier frequency,  $ x^k $  is the exponentially   distributed channel power with unit mean for the tagged RBS, $ d_{i,j} $ is the distance between user $ i $ and BS $ j $, and $ \alpha=2.5 $  is the path-loss exponent. The path-gain for THz communication is calculated by $ h_{i,j}^k= \left({c}/{4\pi f^k_{\mathrm{THz}}}\right)^2 d^{-2}_{i,j} e^{-a(f^k_{\mathrm{THz}})d_{i,j}} $, where $ f^k_{\mathrm{THz}} $ is the operating THz frequency, $ e^{-a(f^k_{\mathrm{THz}})d_{i,j}} $ is the path-loss caused by the molecular absorption \cite{multiband}, and
$ a(f^k_{\mathrm{THz}})=0.0033 $ is a molecular absorption coefficient. The operating frequency of the $ k $th subchannel is  $ f^k_{\mathrm{THz}}=f^c_{\mathrm{THz}}+\dfrac{W}{{C}}\left(k-1-\dfrac{{C}-1}{2}\right)$, where $ W $ is total bandwidth of THz frequency, $ C $ is the total number of subchannels in THz frequency, and $ f_c=340\mathrm{GHz} $ is the central THz frequency  \cite{THz-PathGain}.  

It is assumed that each FeMBB user achieves its QoS requirement by allocating a subchannel. In addition, each subchannel contains $M=7$ mini-slots, and each eURLLC user achieves its QoS requirement by allocating one of these mini-slots.
We consider $10$ FeMBB users with minimim data rate requirement of $R^{\mathrm{min}} =1\mathrm{Mbps} $ and $10$ eURLLC users with maximum decoding error probability of $\epsilon= 10^{-5} $.  The white noise at each subchannel is $ -174 \mathrm{dB} $. Furthermore, the other simulation parameters are $ L_B=100\mathrm{symbols} $, $ D=60 \mathrm{bits}$, and $ T=0.5 \mathrm{ms}$. 
For the implementation of DRL methods, we use PyTorch 1.8.1 with Python 3.7. 

\subsection{Numerical Results and Discussions}
\begin{figure}
	\centering
	\includegraphics[width=0.45\textwidth]{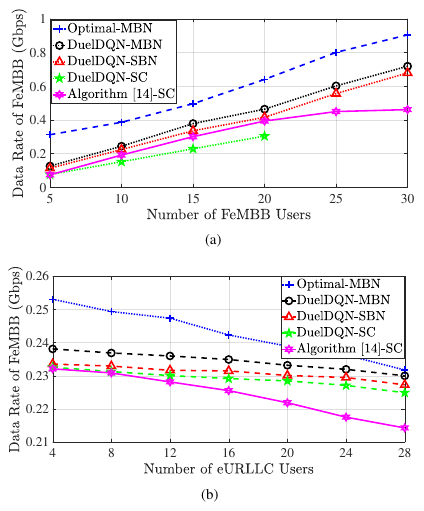}
	\caption{(a) Total data rate of FeMBB users vs. number of FeMBB users, and (b) total data rate of FeMBB users vs. number  of eURLLC users.}
	\label{fig:simulation_2}
\end{figure}

 \CB 

\subsubsection{Efficacy of DuelDQN Method} In Fig. \ref{fig:simulation_1}, we compare the convergence rate and performance of   DQN, double DQN, and DuelDQN methods  for JNSA problem in MBN.  As seen from Fig.~\ref{fig:simulation_1}\CL a\CB,  DuelDQN converges faster than the others. Moreover, Fig. \ref{fig:simulation_1}\CL b \CB illustrates that DuelDQN achieves almost the same performance as DQN with a higher convergence rate and the performance loss compared to the optimal solution is minimal. The optimal solution is obtained by the Mosek toolbox \cite{Mosek}.  

\subsubsection{Efficacy of MBN}
In Figs. \ref{fig:simulation_2}\CL a \CB and \ref{fig:simulation_2}\CL b\CB,  we compare our proposed DuelDQN method  for JNSA problem in MBN (DuelDQN-MBN) and in single-band networks (DuelDQN-SBN)  with the optimal solution (Optimal-MBN). We also employ the DuelDQN method  for subchannel allocation problem in a single-cell network (DuelDQN-SC), and compare it with the algorithm proposed in \cite{TCOMM-2020} which was originally proposed for subchannel allocation in a single-cell network, wherein one RBS serves FeMBB and eURLLC terrestrial users and for FeMBB users, no QoS is guaranteed. 

Fig. \ref{fig:simulation_2}\CL a \CB  shows that, as the number of FeMBB users increases, the total data rate increases since more users have to reach their minimum data rate. Besides,  DuelDQN-MBN obtains a near-optimal performance  compared to the benchmarks. It should be noted that since there are 20 subchannels, each of which should only be allocated to one FeMBB user, the DuelDQN-SC becomes infeasible where the number of FeMBB users is more than 20. Moreover, since in \cite{TCOMM-2020} no QoS constraint is considered for FeMBB users, the algorithm proposed in \cite{TCOMM-2020} obtains a higher data rate in comparison with the DuelDQN-SC.

Fig. \ref{fig:simulation_2}\CL b \CB shows that with the increase in the number of eURLLC users, the total data rate of FeMBB users decreases, due to the increase in the number of mini-slots allocated to  eURLLC users and puncturing of FeMBB users.  

From Figs. \ref{fig:simulation_2}\CL a \CB and \ref{fig:simulation_2}\CL b\CB, it is observed that DuelDQN-MBN due to the THz communication leads to a higher data rate in comparison with DuelDQN-SBN  and DuelDQN-SC.  

\subsubsection{Robustness of DuelDQN Method}
\begin{figure}
	\centering
	\includegraphics[width=0.45\textwidth]{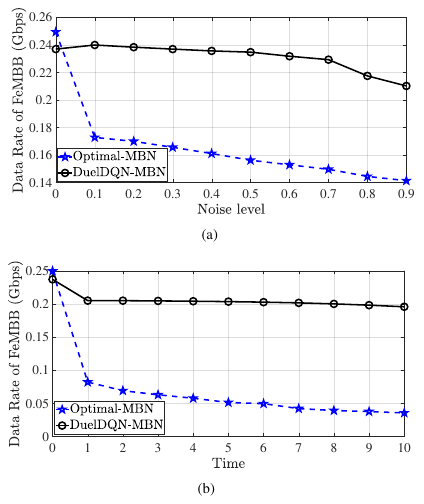}
	\caption{ (a) Total data rate of FeMBB users vs. noise level in CSI, and (d) total data rate of FeMBB users vs. users' mobility time.}
	\label{fig:simulation_3}
\end{figure}
Fig. \ref{fig:simulation_3}\CL a \CB illustrates the impact of CSI error   on the performance of DuelDQN and optimal optimization solution.   To do so, similar to \cite{robustness}, in the first stage, we obtain the user association and subchannel allocation, then we add noise to the path-gains as $ h_{i,j}^k=\sqrt{\delta}h_{i,j}^k+\sqrt{\delta-1} \omega$, in which $ \delta $ is the noise level and $  \omega $ is an independent and identically distributed value with $ CN(0,1) $. In Fig. \ref{fig:simulation_3}\CL a\CB, it can be seen that, because the optimal solution relies on perfect CSI, the reduction in data rate is more severe than in DuelDQN. The DuelDQN approach is thus more robust to changes in network conditions.

Moreover, in Fig.  \ref{fig:simulation_3}\CL b\CB, we investigate the impact of users' mobility on the performance of the optimal and DuelDQN approaches. For generating this figure, at the first step, we obtain the user association and subchannel allocation; afterward, we assume all users move away from their assigned BS with the speed of $ 2\mathrm{m/s} $.  From Fig. \ref{fig:simulation_3}\CL b\CB, it can be observed that the data rate reduction in optimal optimization solution is more severe than that of DuelDQN approach, and DuelDQN is robust when users are mobile. 

\section{CONCLUSION}
The ever increasing  QoS, QoE, and QoPE requirements for future 6G wireless applications are envisioned with new KPIs and  KETs. We described how the novel KETs bring many new resource management challenges, individually or collectively, followed by a discussion on the  general and specific challenges to each KET, and from the cross-layer design and ML-enabled systems perspectives. A  method of network selection and subchannel allocation based on DRL has been proposed for a 6G multi-band wireless network with coexisting FeMBB and eURLLC services.

\end{document}